\documentclass[a4paper, 10pt, conference]{IEEEtran}
\IEEEoverridecommandlockouts
\usepackage{cite}
\usepackage{amsmath,amssymb,amsfonts}
\usepackage{algorithmic}
\usepackage{graphicx}
\usepackage{textcomp}
\usepackage{xcolor}
\def\BibTeX{{\rm B\kern-.05em{\sc i\kern-.025em b}\kern-.08em
    T\kern-.1667em\lower.7ex\hbox{E}\kern-.125emX}}

\usepackage{mathtools}
\usepackage{url}

\usepackage{listings}
\usepackage{tikz}
\usepackage{environ}
\usepackage{pgfplots}
\usepackage{soul}

\begin{document}

\title{Curb Your Self-Modifying Code*\\
\thanks{This work was supported by the Hasler Foundation under grant No.\ 21017.}
}

\author{\IEEEauthorblockN{Patrik Christen}
\IEEEauthorblockA{\textit{Institute for Information Systems} \\
\textit{FHNW}\\
Olten, Switzerland \\
patrik.christen@fhnw.ch}
}

\maketitle

\begin{abstract}
Self-modifying code has many intriguing applications in a broad range of fields including software security, artificial general intelligence, and open-ended evolution. Having control over self-modifying code, however, is still an open challenge since it is a balancing act between providing as much freedom as possible so as not to limit possible solutions, while at the same time imposing restriction to avoid security issues and invalid code or solutions. In the present study, I provide a prototype implementation of how one might curb self-modifying code by introducing control mechanisms for code modifications within specific regions and for specific transitions between code and data. I show that this is possible to achieve with the so-called allagmatic method -- a framework to formalise, model, implement, and interpret complex systems inspired by Gilbert Simondon's philosophy of individuation and Alfred North Whitehead's philosophy of organism. Thereby, the allagmatic method serves as guidance for self-modification based on concepts defined in a metaphysical framework. I conclude that the allagmatic method seems to be a suitable framework for control mechanisms in self-modifying code and that there are intriguing analogies between the presented control mechanisms and gene regulation.
\end{abstract}

\begin{IEEEkeywords}
allagmatic method, code-data duality, self-modifying code, gene regulation
\end{IEEEkeywords}

\section{Introduction}

A robot twin of the president of the United States is instructed to initialise its shutdown sequence. He (or it?) responses unmoved by saying ``I wrote that part of my code out''. The scientist woman that created him answers exhilarated with ``You rewrote your own code? You've reached true AI''. Of course this dialogue is from a TV series, Inside Job, and of course the robot tries to destroy humanity shortly afterwards. Nevertheless, it illustrates the security issue of loosing control over code that can rewrite itself. It furthermore links true artificial intelligence or artificial general intelligence to the capability of modifying its own code.

The steps involved transforming the robot president from an AI to a true AI are not explained and thus the scenario remains science fiction. The behaviour of self-modifying code, however, is hard to predict and control, so that the TV series has a point. \textit{Control} in this context means to determine where in the program code modifications are possible and what kind of modifications are possible. Programmers have exploited the characteristics of self-modifying code for both protecting and attacking software. E.g., Shan and Emmanuel \cite{Shan.2011} proposed an obfuscation algorithm based on self-modifying code to prevent mobile agent code from attacks. They showed that obfuscation that is the deliberate creation of difficult understandable code, provides effective protection against reverse engineering sensitive information within mobile agent code. Mi et al. \cite{Mi.2015} provide an overview of methods to protect software intellectual property from reverse attackers. In these cases, self-modifying code is used to make code harder to understand, which costs attackers more time and requires more advanced programming skills. Self-modifying code has also been used to attack software, e.g., by hiding internal logic of a virus making it hard for antivirus software to detect it \cite{Anckaert.2007}.

As alluded to in the TV series dialogue above, self-modifying code plays a role in the field of artificial general intelligence. It has been argued that an artificial general intelligence is likely to possess the ability of self-reflection, which includes making modifications to itself at runtime \cite{Steunebrink.2012}. This is assumed because it seems unlikely to create a completely predetermined program that satisfies general intelligence conditions without requiring adaptation at some later point in time \cite{Steunebrink.2012}. E.g., Gödel machines can completely rewrite their own code provided an embedded proof searcher can prove that the rewrite is useful \cite{Schmidhuber.2003}. The idea is that improvements are implemented through self-modifying code imposing no limits on possible solutions as this would be the case with hardwired algorithms.

Similarly, the field of open-ended evolution \cite{Packard.2019.I,Packard.2019.II,Stanley.2017} motivates the use of self-modifying code by imposing as few limits as possible to novelties that emerge in a system. Banzhaf et al. \cite{Banzhaf.2016} proposed three types of novelties: variation, innovation, and emergence. Variation is a novelty within the model of a system and thus a change to a model instance is required to capture it, innovation is a novelty that requires changes of the model of a system to capture it, and emergence is a novelty that requires changes to the metamodel of a system \cite{Banzhaf.2016} to capture it.

The first type of novelty, variation, does not require self-modifying code as it can be implemented with model parameters. The other two types of novelty, innovation and emergence, are not as straightforward to implement. Some innovations might be implemented with generic programming, e.g., generics and templates providing types as parameters \cite{Czarnecki.2000}. Self-modifying code, however, is a useful approach in cases where such programming concepts are difficult to apply or missing.

Christen \cite{Christen.2021a} recently provided a self-modifying code prototype and a formalism how it could be applied within the so-called allagmatic method \cite{Christen.2020a,Christen.2019} -- a framework to formalise, model, implement, and interpret complex systems. Thereby, the allagmatic method serves as guidance for self-modification based on concepts defined in a metaphysical framework. Within this framework, it implements philosophical concepts such as structure, operation, and control.

Implementing philosophical concepts allows definition and interpretation of the model and code. It is therefore helpful if one is interested in what is happening in the code itself. E.g., to identify and categorise novelties in the field of open-ended evolution and, as it is suggested here, to get a handle on changes in program code due to self-modification. However, if one uses low-level programming languages, which are a straightforward way to implement self-modifying code \cite{Banzhaf.2016}, definition and interpretation become hard if not impossible. The instruction set in these languages is limited and close to machine code, which makes them in general difficult to relate to higher level concepts. If used to implement self-modifying code, the structure and behaviour of the code needs to be inferred somehow, e.g., like in a biological experiment. To avoid that and to better explain and control changes occurring in the code, I propose here to use the allagmatic method and the high-level language C\#.

Already John von Neumann was interested in self-modification in the context of self-reproducing automata \cite{VonNeumann.1966,Burks.1970}. He originally conceived of cellular automata as abstract models of machines capable of self-reproduction and was able to construct a self-reproducing universal Turing machine embedded within a twenty-nine-state two-dimensional cellular automaton \cite{Ilachinski.2001}. What is striking and relevant in the context of self-modifying code, is that von Neumann not only introduced a machine blueprint that is copied during self-reproduction but also that the data of this blueprint can be viewed as consisting of active instructions to be executed and as an assemblage of passive information that must be copied and attached to the offspring machine \cite{Ilachinski.2001}. As we will see in the following, the present approach to self-modifying code used the same duality between data as active instructions and as passive information.

Self-modifying code seems to be related to deep concepts and a promising approach to tackle the problems mentioned at the beginning. These studies, however, also show that controlling self-modifying code is an issue that needs some attention. It is a balancing act between providing as much freedom as possible to not limit possible solutions and restriction to avoid security issues and invalid code or solutions. In the present study, I provide a prototype implementation of how one might curb self-modifying code by introducing control mechanisms for code modifications within specific regions (section Partial Code Modification) and for specific transitions between code and data (section Code and Data Duality). The allagmatic method thereby serves as a framework for defining regions of code modifications and transitions between code and data. It is described in the next section The Allagmatic Method, followed by the sections Partial Code Modification and Code and Data Duality, describing a prototype implementation of the presented control mechanisms for self-modifying code.

\section{The Allagmatic Method}

\begin{figure}[tbp]
\centerline{\includegraphics[width=1.0\columnwidth]{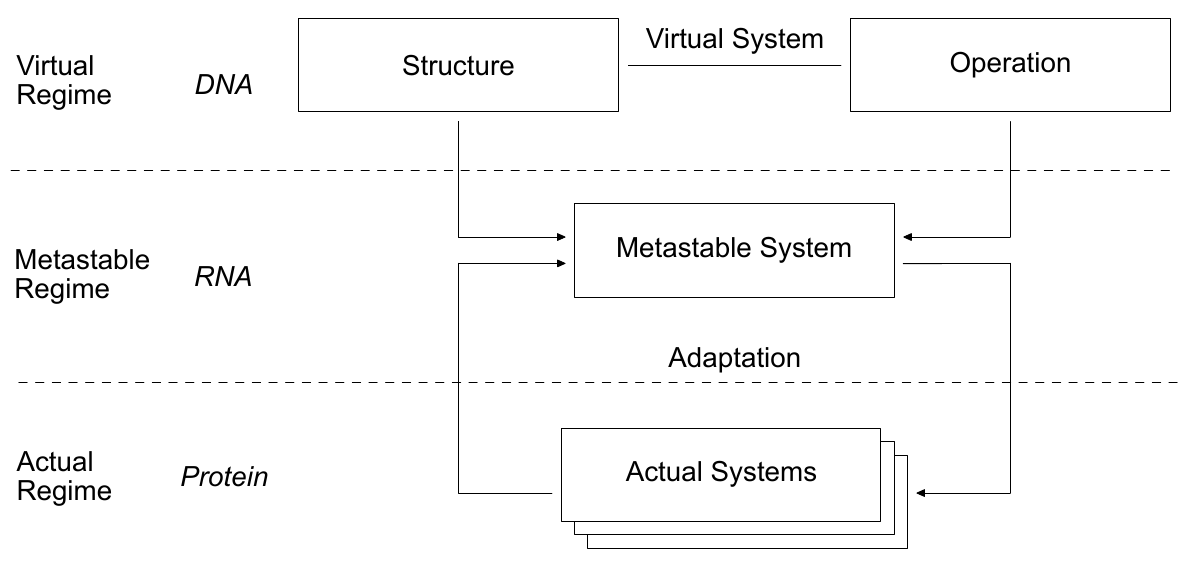}}
\caption{Schematic overview of the allagmatic method. In the virtual regime, a system is described in a most abstract way in terms of general structures and operations. In the metastable regime, concretisation of the system is described. In the actual regime, the fully concretised system is executed. Adaptation usually occurs between the metastable and actual regime since only after running the system it can provide feedback that can be accounted for in the metastable regime. An analogy is made to gene regulation, where the flow of genetic information is from DNA to RNA to protein \cite{Alberts.2002}. Figure adapted from \cite{Christen.2019}.}
\label{fig1}
\end{figure}

Philosophy has a long history and strong tradition for describing and categorising anything \cite{Thomasson.2019}. It therefore provides a rich source for inspiration and guidance, especially in the field of complex systems modelling. The allagmatic method \cite{Christen.2020a,Christen.2019} implements philosophical concepts by defining them abstractly in a \textit{virtual} regime, making them more concrete by creating concrete model instances in the \textit{metastable} regime, and executing these concrete instances in the \textit{actual} regime (Fig.~\ref{fig1}). \textit{Metastability thereby refers to the potentiality of a system to change} \cite{DelFabbro.2021,Simondon.2020}. It consists of a system metamodel describing philosophical concepts with object-oriented programming. Step-wise concretisation of model instances with parameters is implemented with generic programming \cite{Czarnecki.2000}. The approach has been successfully applied to automatic programming, where it guided the definition of code building blocks to be automatically implemented in cellular automaton and artificial neural network models \cite{Christen.2021b}. It has been also used to define and interpret control and adaptation in a complex system \cite{DelFabbro.2020}.

The allagmatic method is inspired by Gilbert Simondon's philosophy of individuation \cite{DelFabbro.2021,Simondon.2020} and Alfred North Whitehead's philosophy of organism \cite{Debaise.2017,Whitehead.1978}. According to Simondon, systems are formed by and consist of intertwined structures and operations that are more and more concretised over time (individuation process) \cite{DelFabbro.2021}. According to him, systems can thus be described from a structural and operational perspective. The system metamodel consists of basic and abstract definitions of structures and operations forming together a (complex) system borrowing Simondon's concepts of structure and operation \cite{Christen.2020a,Christen.2019}. This is extended with the concepts of entity, control, and adaptation borrowed from Whitehead \cite{DelFabbro.2020}. The system metamodel therefore consists of operations and related structures defining these concepts \cite{DelFabbro.2020,Christen.2020a}.

In the most abstract sense and thus in the virtual regime, a model of a system (system metamodel) $\mathcal{SM}$ is structurally described with a network of interconnected \textit{entities} $\mathcal{E}$ that are in a certain \textit{state} $\hat{e}_i$ of a set of possible states $Q$. The so-called \textit{milieu} $\hat{\mathcal{M}}_i$ of an entity describes its environment regarding interactions and thus the entities connected to it. From an operational perspective, a model of a system can be described with an \textit{update function} $\phi$ that changes the states of entities according to some predefined \textit{update rules} $\mathcal{U}$ taking into account the states of entities in the milieu of an entity in an iterative process. An \textit{adaptation function} $\psi$ furthermore changes the update function or more precisely the rules $\mathcal{U}$ it contains. These are the basic building blocks which allow to describe systems from the point of view of complex systems. They are abstractly defined in the system metamodel and implemented as a class. Models are concretised by making these building blocks more and more concrete, e.g., by specifying the number of entities, their state type, and specific update rules. A more detailed and formal description of the allagmatic method can be found here \cite{Christen.2020a}.

In addition, I make an analogy to molecular biology, i.e., gene regulation, providing a more concrete description of the allagmatic method. One of the most fundamental processes in molecular biology is the synthesis of proteins based on information in DNA. DNA thereby does not direct the creation of proteins itself, but instead uses RNA as an intermediary molecule \cite{Alberts.2002}. In a process called transcription, the appropriate sequence of nucleotides in the DNA is first copied into RNA \cite{Alberts.2002}. These RNA copies of segments of the DNA are then used directly as templates to direct protein synthesis in a process called translation \cite{Alberts.2002}. RNA transcripts are affected by a series of processing steps before they are translated into proteins, which determines the concrete protein blueprint \cite{Alberts.2002}. The analogy to the allagmatic method is depicted in Fig.~\ref{fig1}: DNA provides some basic building blocks, as do classes that describe the system in the virtual regime. In gene regulation, the final product is a protein ready to perform a certain task. In the allagmatic method, the final product is a model ready to be executed, which is described by the system in the actual regime. Finally, there are some intermediate steps between the basic building blocks DNA and protein, which are based on RNA. This is described by the system in the metastable regime in the allagmatic method.

In the next two sections, the allagmatic method is used as a framework and as guidance to restrict self-modifying code to specific regions and to define specific transitions between code and data.

\section{Partial Code Modification}

The system metamodel of the allagmatic method describes a system in general terms and based on philosophical concepts that have proved to be useful in describing a broad range of real-world systems. With that, it imposes little limitations on possible solutions. It can thus be used as a minimal restriction for self-modifying code without limiting its creative potential. Additionally, not all parts of the system metamodel require self-modifying code. There are also possible restrictions that can be predetermined and imposed by the specific application. They are not considered in the following prototype implementation in C\#.

The main program is written in the class \texttt{Program} and an additional class \texttt{Entity} describes an entity. An entity object $\hat{e}_i\in Q$ has a field representing its state of set $Q$ (or more than one field representing multiple states of multiple sets), its update rules $\mathcal{U}$, and milieu $\hat{\mathcal{M}}_i$. Class fields therefore describe structures. Operations are described with methods in the same class. Specifically, an \texttt{UpdateFunction} representing $\phi$ in the system metamodel. It iteratively updates the states of an entity. In the main program, a list of entity objects is created representing the entities tuple $\mathcal{E}$. The \texttt{List} class \cite{ListClass} provided by the .NET API is used. It is a dynamic list allowing to change its size at runtime. The number of entities is a model parameter concretised in the metastable regime. Generic programming \cite{Czarnecki.2000} allows now to implement some parts of this without any self-modifying code. The field entity states (a list of one or more states) is implemented with a dynamic type \cite{Skeet.2019}. This allows the abstract definition of entity states in the virtual regime and the concretisation of it through concrete value assignment in the metastable regime. Methods (operations) working on or with entity states (structure), use pattern matching \cite{Skeet.2019} to reveal concrete states if necessary. The milieu of an entity, that is the neighbours or linked entities, is described with an entity object field representing $\hat{\mathcal{M}}_i$ in $\mathcal{M}$. It is a dynamic list again containing references to the linked entities. No self-modifying code is necessary here. Every possible network structure of linked entities can be configured in these milieus without any restriction. This is different for the update function $\phi$. There are an infinite number of possibilities how to update entity states and which neighbouring states to take into account. The logic of the update function or update rules are described with the structure $\mathcal{U}$. It is here implemented as a string containing code that implements the update function. This is code as data and will be discussed in more detail in the next section. 

The important thing about partial code modification is that it can be used to define update rules $\mathcal{U}$ as self-modifiable. Within $\mathcal{U}$, whatever is implemented, it takes as an input the entity states of the $i$-th milieu entities, $\hat{\mathcal{M}}_i^{(\bar{t})}$, of the current iteration, $\bar{t}$, and it outputs the entity state of the $i$-th entity, $\hat{e}_i^{(\bar{t}+1)}$, of the next iteration, $\bar{t}+1$. This is used to guide the generation of code with the self-modifying code prototype developed earlier \cite{Christen.2021a}. The entity state type reveals possible operations. In addition, the milieu of an entity is known at the time of updating an entity's state. These states of neighbouring entities are used as guidance to generate code. The update function $\phi$ is implemented as a method that compiles and runs the code in $\mathcal{U}$. It is described in more detail in the next section.

\section{Code-Data Duality}

Program code that is capable of modifying itself, i.e., self-modifying code, can be achieved by introducing a duality between code and data. It means that code or some parts of the code are not only program code for execution, but also accessible as data. In the present study, a self-modifying code prototype implemented earlier is used \cite{Christen.2021a}. It is integrated into the allagmatic method, and the partial code modification prototype described in the previous section. Self-modification is restricted to the update rules $\mathcal{U}$ of the update function $\phi$. The update function of an entity iteratively changes the state of an entity (or states of an entity) according to some update rules. Code-data duality is implemented by storing the update rules $\mathcal{U}$ in a string variable. $\mathcal{U}$ can be regarded as the logic of the update function in the form of data. It follows that $\mathcal{U}$ as code is the update function $\phi$, which is code to be executed. This is implemented with a method \texttt{UpdateFunction} that reads the update rules as data (it reads its own logic as data), compiles and runs them.

The allagmatic method models complex systems as a network of entities that evolves iteratively. Thereby, an entity changes its state(s) based on its current state $\hat{e}_i$ and the states of its neighbouring entities stored in the milieu $\hat{\mathcal{M}}_i$ of this entity $\hat{e}_i$. Neighbours therefore means the entities connected to a particular entity. A state update in a single iteration is performed by invoking the update function $\phi$, which executes the update rules described in $\mathcal{U}$. These rules are described in the form of program code stored in a string and thus in the form of code as data. Entity states have to transition from the currently running program of the simulation into the update rules $\mathcal{U}$, therefore from code to data, and after the data is run as code, they have to transition back from the data to code (to the running simulation) in the form of updated entity states.

The first transition, that is code to data, is implemented by replacing variables as strings in the data with the values of the same variables in the code (string interpolation \cite{Stuart.2013}), e.g., the variable \texttt{entityState} is declared as a string \texttt{"dynamic entityState"} in $\mathcal{U}$ as data and then replaced by the value of \texttt{entityState} of the running simulation program and thus as code. This replacement can be achieved with \texttt{updateRules.Replace("entityState", entityState.ToString())}, where \texttt{updateRules} is a string representing $\mathcal{U}$, \texttt{"entityState"} is a string representing the variable \texttt{entityState} as data, and \texttt{entityState} is a variable representing the variable \texttt{entityState} as code. The \texttt{ToString} method \cite{ToStringMethod} creates a string out of the value of the variable \texttt{entityState}. The \texttt{Replace} method \cite{ReplaceMethod} replaces all the substrings \texttt{"entityState"} of the string \texttt{updateRules} with the value of \texttt{entityState} as a string. It is also possible to directly perform string interpolation in C\# with the special character \$, which identifies a string as an interpolated string \cite{InterpolatedStrings}. The newly generated string \texttt{updateRules} is then sent to the method \texttt{UpdateFunction}, where it is treated as code. It is compiled and then executed using the previously developed self-modifying code prototype \cite{Christen.2021a}. In short, it converts the string \texttt{updateRules} into a syntax tree object \cite{CSharpSyntaxTreeClass} and this tree is then compiled as an object without writing out a source file. This compilation object can be executed at runtime. 

Now the second transition needs to happen to get the result of the update function $\phi$, that is the updated entity states. It is a transition back from data to code. It is implemented here by redirecting the console output stream of the data executed as code to a variable in the code of the running simulation program. Before the data is executed as code, the output stream is set to a \texttt{StringWriter} object \cite{StringWriterClass}. During execution of the data as code, the variable \texttt{entityState} is then printed out in the console as usual with \texttt{Console.WriteLine(entityState)} \cite{WriteLineMethod}. It is, however, not printed out into the console, it is stored in the \texttt{StringWriter} object due to the redirection of the output stream. From there, the updated \texttt{entityState} variable can be used in the running simulation program. The transition from data back to code is therefore complete.

\section{Discussion and Conclusion}

Self-modifying code is hard to understand and control, and its behaviour is hard to predict. It has nevertheless intriguing applications in fields such as software security \cite{Mi.2015,Shan.2011}, artificial general intelligence \cite{Steunebrink.2012,Schmidhuber.2003}, and open-ended evolution \cite{Christen.2021a,Banzhaf.2016}. The present study addresses the issue of controlling self-modifying code by introducing control mechanisms for partial code modifications and specific transitions between code and data. Code regions and specific transitions between code and data are defined with the help of the allagmatic method \cite{Christen.2020a,Christen.2019}. It serves as a framework to curb self-modifying code. The study also provides a prototype implementation of the control mechanisms based on an earlier developed self-modifying code prototype \cite{Christen.2021a}. 

The provided prototype is implemented in C\#, however, other high-level programming languages might be used as well, e.g., automatic programming of code as a string has been shown in C++ by writing the code string in a source file, compiling and running it with the \texttt{system} method \cite{Christen.2021b}. Dynamic types can be implemented with C++ templates as type parameters \cite{Czarnecki.2000}. Transitions between code and data might be defined via \texttt{main} method parameters and return values. It has to be noted that writing out into a source file and running the compiler in this way in C++ is much more time consuming than compiling and running an object in C\#.

Additionally, C\# has been used here because it allows rapid prototyping, and provides a vast API for implementing higher level concepts such as reflection. However, there are programming languages that provide useful features to implement self-modifying code. E.g., the programming language Push is designed to implement evolutionary computation systems and provides a data type to represent program code \cite{Spector.2002}. Similarly, in the programming language Lisp, program code is made of lists, its primary data structure allowing programs to modify their own code \cite{McCarthy.1978}.  Although in such programming languages it is straightforward to transition from code to data and back again, C\# is used here to make use of its API and familiarity with its C++ like syntax and programming paradigms.

It is interesting to mention that the self-modifying code prototype developed earlier \cite{Christen.2021a} imposes restrictions that do not necessarily limit creativity but help increase software security. Code is generated from a finite list of predefined words, which provides the building blocks to program operations of virtually any kind. However, the code is limited with respect to data access and writing. It can only operate on predefined variables, i.e., entity states, and it can only create new identifiers with the naming \texttt{identifierX}, where \texttt{X} is replaced by an integer value counting newly defined identifiers. It is thus not possible to inject malicious code via identifier names.

The duality between code and data seems to be directly comparable to the duality between operation and structure in Simondon's philosophy of individuation \cite{DelFabbro.2021,Simondon.2020}. According to Simondon, an object as part of a system has a structure and acts according to an operation. Structure and operation are intertwined forming a system, e.g., cooling fins as part of a larger system, an air-cooled engine in this example, have a certain structure or geometry and at the same time fulfil the operations of cooling and structural stabilisation. Similarly, code as data is structure and data as code is operation of a system. In the allagmatic method, this is, for example, the update rule $U$, which is code as data, and the update function $\phi$, which is data as code since it compiles and runs $U$. Program code can thus be described structurally via data and operationally via code.

The analogy between gene regulation and the allagmatic method can now be further developed. I present here two control mechanisms for partial code modification and specific transitions between code and data. Partial code modification is achieved by limiting the modifications to certain parts of the allagmatic method. In the present prototype, the structure $\mathcal{U}$ is modifiable. It describes the logic of the update function or dynamics of the system. The program can thus only modify $\mathcal{U}$. This is happening in the virtual regime of the allagmatic method and is comparable to the selection of or limitation to a specific nucleotides sequence in DNA. Another restriction imposed by the allagmatic method in the virtual regime is the input and output of the update function $\phi$, which are both entity states of the same (data) type. The type is concretised as soon as we transition to the metastable regime. DNA transcription imposes similar restrictions. Both, DNA and RNA are based on sequences of nucleotides \cite{Alberts.2002}. They are of the same type, and thus also here the input and output are defined. Once RNA is transcribed, it is modified in several processing steps before it is used as a template to synthesise protein \cite{Alberts.2002}. It is a messy place, where the RNA can be changed significantly \cite{Alberts.2002}. RNA splicing is one such process. It modifies newly transcribed RNA (called pre-RNA) into mature RNA by splicing-out certain sequences (introns) \cite{Clancy.2008,WilliamRoy.2006,Gilbert.1978}. It thereby brings RNA in a form that is more suitable to translate into protein. In the allagmatic method, such processes are described as operations and implemented as methods in the virtual regime. They are run in the metastable regime, where they concretise the virtual system into a metastable and finally actual system. These are therefore specific transitions that are controlled or defined by the allagmatic method. RNA splicing is thus analogous to the allagmatic method in terms of both, the specificity of transitions and maybe even more remarkably, the concretisation process. Finally, the actual regime describes the fully concretised system that can be executed or is ready to perform a certain task. This is analogous to a protein that can also perform a function.

I conclude that the allagmatic method serves as guidance to implement control mechanisms for defining code modifications for specific code parts and concepts as well as for defining specific transitions between code and data. Furthermore, since the philosophy that inspired the allagmatic method is heavily influenced by biological research, the allagmatic method seems to be well suited to make analogies to biology and thus provides a framework to describe and explore complex biological systems.

\section*{Acknowledgment}
%
I thank the Complexity Club for valuable input on an earlier version of the manuscript.

\bibliographystyle{ieeetr}
\bibliography{Self_Modifying_Code}


\end{document}